\documentclass[twocolumn,aps,prc,showpacs,superscriptaddress,amsmath,amssymb,floatfix]{revtex4}

\usepackage{epsfig}
\usepackage{amssymb}

\usepackage{graphicx}
\usepackage[figuresright]{rotating}

\newcommand{\agev}{\mbox{$A$~GeV}}               
\newcommand{\gevc}{\mbox{GeV$/c$}}
\newcommand{\gevcc}{\mbox{GeV$/c^2$}}

\newcommand{\gevf}{\mbox{GeV/fm$^3$}}
\newcommand{\fmm}{\mbox{fm$^{-3}$}}

\newcommand{\fmc}{\mbox{fm$/c$}}

\newcommand{\rb}[1]{\mbox{\textrm{\scriptsize #1}}}

\newcommand{\ket}{\ensuremath{KE_{\rb{T}}}}
\newcommand{\pt}{\ensuremath{p_{\rb{T}}}}
\newcommand{\mt}{\ensuremath{m_{\rb{T}}}}
\newcommand{\vi}{\ensuremath{v_{\rb{2}}}}
\newcommand{\phiep}{\ensuremath{\Phi_{\rb{2EP}}}}

\begin{document}
\title{Elliptic flow of $\Lambda$ hyperons in Pb+Pb collisions at
158\agev}





\affiliation{NIKHEF, Amsterdam, Netherlands.}
\affiliation{Department of Physics, University of Athens, Athens,
Greece.} \affiliation{Comenius University, Bratislava, Slovakia.}
\affiliation{KFKI Research Institute for Particle and Nuclear
Physics,
             Budapest, Hungary.}
\affiliation{MIT, Cambridge, USA.} \affiliation{The
H.Niewodniczanski Institute of Nuclear Physics, Polish Academy of
Sciences, Cracow, Poland.} \affiliation{Gesellschaft f\"{u}r
Schwerionenforschung (GSI),
             Darmstadt, Germany.}
\affiliation{Joint Institute for Nuclear Research, Dubna, Russia.}
\affiliation{Fachbereich Physik der Universit\"{a}t, Frankfurt,
Germany.} \affiliation{CERN, Geneva, Switzerland.}
\affiliation{University of Houston, Houston, TX, USA.}
\affiliation{Institute of Physics \'Swi\c etokrzyska Academy,
Kielce, Poland.} \affiliation{Fachbereich Physik der
Universit\"{a}t, Marburg, Germany.}
\affiliation{Max-Planck-Institut f\"{u}r Physik, Munich, Germany.}
\affiliation{Institute of Particle and Nuclear Physics, Charles
             University, Prague, Czech Republic.}
\affiliation{Department of Physics, Pusan National University,
Pusan,
             Republic of Korea.}
\affiliation{Nuclear Physics Laboratory, University of Washington,
             Seattle, WA, USA.}
\affiliation{Atomic Physics Department, Sofia University
St.~Kliment
             Ohridski, Sofia, Bulgaria.}
\affiliation{Institute for Nuclear Studies, Warsaw, Poland.}
\affiliation{Institute for Experimental Physics, University of
Warsaw,
             Warsaw, Poland.}
\affiliation{Rudjer Boskovic Institute, Zagreb, Croatia.}
\affiliation{Warsaw University of Technology, Warsaw, Poland.}
\affiliation{Department of Chemistry, SUNY Stony Brook, USA.}

\author{C.~Alt}
\affiliation{Fachbereich Physik der Universit\"{a}t, Frankfurt,
Germany.}
\author{T.~Anticic}
\affiliation{Rudjer Boskovic Institute, Zagreb, Croatia.}
\author{B.~Baatar}
\affiliation{Joint Institute for Nuclear Research, Dubna, Russia.}
\author{D.~Barna}
\affiliation{KFKI Research Institute for Particle and Nuclear
Physics,
             Budapest, Hungary.}
\author{J.~Bartke}
\affiliation{The H.Niewodniczanski Institute of Nuclear Physics,
Polish Academy of Sciences, Cracow, Poland.}
\author{L.~Betev}
\affiliation{CERN, Geneva, Switzerland.}
\author{H.~Bia{\l}\-kowska}
\affiliation{Institute for Nuclear Studies, Warsaw, Poland.}
\author{C.~Blume}
\affiliation{Fachbereich Physik der Universit\"{a}t, Frankfurt,
Germany.}
\author{B.~Boimska}
\affiliation{Institute for Nuclear Studies, Warsaw, Poland.}
\author{M.~Botje}
\affiliation{NIKHEF, Amsterdam, Netherlands.}
\author{J.~Bracinik}
\affiliation{Comenius University, Bratislava, Slovakia.}
\author{R.~Bramm}
\affiliation{Fachbereich Physik der Universit\"{a}t, Frankfurt,
Germany.}
\author{P.~Bun\v{c}i\'{c}}
\affiliation{CERN, Geneva, Switzerland.}
\author{V.~Cerny}
\affiliation{Comenius University, Bratislava, Slovakia.}
\author{P.~Christakoglou}
\affiliation{Department of Physics, University of Athens, Athens,
Greece.}
\author{P.~Chung}
\affiliation{Department of Chemistry, SUNY Stony Brook, USA.}
\author{O.~Chvala}
\affiliation{Institute of Particle and Nuclear Physics, Charles
             University, Prague, Czech Republic.}
\author{J.G.~Cramer}
\affiliation{Nuclear Physics Laboratory, University of Washington,
             Seattle, WA, USA.}
\author{P.~Csat\'{o}}
\affiliation{KFKI Research Institute for Particle and Nuclear
Physics,
             Budapest, Hungary.}
\author{P.~Dinkelaker}
\affiliation{Fachbereich Physik der Universit\"{a}t, Frankfurt,
Germany.}
\author{V.~Eckardt}
\affiliation{Max-Planck-Institut f\"{u}r Physik, Munich, Germany.}
\author{D.~Flierl}
\affiliation{Fachbereich Physik der Universit\"{a}t, Frankfurt,
Germany.}
\author{Z.~Fodor}
\affiliation{KFKI Research Institute for Particle and Nuclear
Physics,
             Budapest, Hungary.}
\author{P.~Foka}
\affiliation{Gesellschaft f\"{u}r Schwerionenforschung (GSI),
             Darmstadt, Germany.}
\author{V.~Friese}
\affiliation{Gesellschaft f\"{u}r Schwerionenforschung (GSI),
             Darmstadt, Germany.}
\author{J.~G\'{a}l}
\affiliation{KFKI Research Institute for Particle and Nuclear
Physics,
             Budapest, Hungary.}
\author{M.~Ga\'zdzicki}
\affiliation{Fachbereich Physik der Universit\"{a}t, Frankfurt,
Germany.} \affiliation{Institute of Physics \'Swi\c etokrzyska
Academy, Kielce, Poland.}
\author{G.~Georgopoulos}
\affiliation{Department of Physics, University of Athens, Athens,
Greece.}
\author{E.~G{\l}adysz}
\affiliation{The H.Niewodniczanski Institute of Nuclear Physics,
Polish Academy of Sciences, Cracow, Poland.}
\author{K.~Grebieszkow}
\affiliation{Warsaw University of Technology, Warsaw, Poland.}
\author{S.~Hegyi}
\affiliation{KFKI Research Institute for Particle and Nuclear
Physics,
             Budapest, Hungary.}
\author{C.~H\"{o}hne}
\affiliation{Fachbereich Physik der Universit\"{a}t, Marburg,
Germany.}
\author{K.~Kadija}
\affiliation{Rudjer Boskovic Institute, Zagreb, Croatia.}
\author{A.~Karev}
\affiliation{Max-Planck-Institut f\"{u}r Physik, Munich, Germany.}
\author{M.~Kliemant}
\affiliation{Fachbereich Physik der Universit\"{a}t, Frankfurt,
Germany.}
\author{S.~Kniege}
\affiliation{Fachbereich Physik der Universit\"{a}t, Frankfurt,
Germany.}
\author{V.I.~Kolesnikov}
\affiliation{Joint Institute for Nuclear Research, Dubna, Russia.}
\author{E.~Kornas}
\affiliation{The H.Niewodniczanski Institute of Nuclear Physics,
Polish Academy of Sciences, Cracow, Poland.}
\author{R.~Korus}
\affiliation{Institute of Physics \'Swi\c etokrzyska Academy,
Kielce, Poland.}
\author{M.~Kowalski}
\affiliation{The H.Niewodniczanski Institute of Nuclear Physics,
Polish Academy of Sciences, Cracow, Poland.}
\author{I.~Kraus}
\affiliation{Gesellschaft f\"{u}r Schwerionenforschung (GSI),
             Darmstadt, Germany.}
\author{M.~Kreps}
\affiliation{Comenius University, Bratislava, Slovakia.}
\author{R.~Lacey}
\affiliation{Department of Chemistry, SUNY Stony Brook, USA.}
\author{A.~Laszlo}
\affiliation{KFKI Research Institute for Particle and Nuclear
Physics,
             Budapest, Hungary.}
\author{M.~van~Leeuwen}
\affiliation{NIKHEF, Amsterdam, Netherlands.}
\author{P.~L\'{e}vai}
\affiliation{KFKI Research Institute for Particle and Nuclear
Physics,
             Budapest, Hungary.}
\author{L.~Litov}
\affiliation{Atomic Physics Department, Sofia University
St.~Kliment
             Ohridski, Sofia, Bulgaria.}
\author{B.~Lungwitz}
\affiliation{Fachbereich Physik der Universit\"{a}t, Frankfurt,
Germany.}
\author{M.~Makariev}
\affiliation{Atomic Physics Department, Sofia University
St.~Kliment
             Ohridski, Sofia, Bulgaria.}
\author{A.I.~Malakhov}
\affiliation{Joint Institute for Nuclear Research, Dubna, Russia.}
\author{M.~Mateev}
\affiliation{Atomic Physics Department, Sofia University
St.~Kliment
             Ohridski, Sofia, Bulgaria.}
\author{G.L.~Melkumov}
\affiliation{Joint Institute for Nuclear Research, Dubna, Russia.}
\author{A.~Mischke}
\affiliation{Gesellschaft f\"{u}r Schwerionenforschung (GSI),
             Darmstadt, Germany.}
\author{M.~Mitrovski}
\affiliation{Fachbereich Physik der Universit\"{a}t, Frankfurt,
Germany.}
\author{J.~Moln\'{a}r}
\affiliation{KFKI Research Institute for Particle and Nuclear
Physics,
             Budapest, Hungary.}
\author{St.~Mr\'owczy\'nski}
\affiliation{Institute of Physics \'Swi\c etokrzyska Academy,
Kielce, Poland.}
\author{V.~Nicolic}
\affiliation{Rudjer Boskovic Institute, Zagreb, Croatia.}
\author{G.~P\'{a}lla}
\affiliation{KFKI Research Institute for Particle and Nuclear
Physics,
             Budapest, Hungary.}
\author{A.D.~Panagiotou}
\affiliation{Department of Physics, University of Athens, Athens,
Greece.}
\author{D.~Panayotov}
\affiliation{Atomic Physics Department, Sofia University
St.~Kliment
             Ohridski, Sofia, Bulgaria.}
\author{A.~Petridis}
\affiliation{Department of Physics, University of Athens, Athens,
Greece.}
\author{M.~Pikna}
\affiliation{Comenius University, Bratislava, Slovakia.}
\author{D.~Prindle}
\affiliation{Nuclear Physics Laboratory, University of Washington,
             Seattle, WA, USA.}
\author{F.~P\"{u}hlhofer}
\affiliation{Fachbereich Physik der Universit\"{a}t, Marburg,
Germany.}
\author{R.~Renfordt}
\affiliation{Fachbereich Physik der Universit\"{a}t, Frankfurt,
Germany.}
\author{C.~Roland}
\affiliation{MIT, Cambridge, USA.}
\author{G.~Roland}
\affiliation{MIT, Cambridge, USA.}
\author{M.~Rybczy\'nski}
\affiliation{Institute of Physics \'Swi\c etokrzyska Academy,
Kielce, Poland.}
\author{A.~Rybicki}
\affiliation{The H.Niewodniczanski Institute of Nuclear Physics,
Polish Academy of Sciences, Cracow, Poland.}
\author{A.~Sandoval}
\affiliation{Gesellschaft f\"{u}r Schwerionenforschung (GSI),
             Darmstadt, Germany.}
\author{N.~Schmitz}
\affiliation{Max-Planck-Institut f\"{u}r Physik, Munich, Germany.}
\author{T.~Schuster}
\affiliation{Fachbereich Physik der Universit\"{a}t, Frankfurt,
Germany.}
\author{P.~Seyboth}
\affiliation{Max-Planck-Institut f\"{u}r Physik, Munich, Germany.}
\author{F.~Sikl\'{e}r}
\affiliation{KFKI Research Institute for Particle and Nuclear
Physics,
             Budapest, Hungary.}
\author{B.~Sitar}
\affiliation{Comenius University, Bratislava, Slovakia.}
\author{E.~Skrzypczak}
\affiliation{Institute for Experimental Physics, University of
Warsaw,
             Warsaw, Poland.}
\author{G.~Stefanek}~\email[Corresponding author. E-mail address: ]{stefanek@pu.kielce.pl}
\affiliation{Institute of Physics \'Swi\c etokrzyska Academy,
Kielce, Poland.}
\author{R.~Stock}
\affiliation{Fachbereich Physik der Universit\"{a}t, Frankfurt,
Germany.}
\author{C.~Strabel}
\affiliation{Fachbereich Physik der Universit\"{a}t, Frankfurt,
Germany.}
\author{H.~Str\"{o}bele}
\affiliation{Fachbereich Physik der Universit\"{a}t, Frankfurt,
Germany.}
\author{T.~Susa}
\affiliation{Rudjer Boskovic Institute, Zagreb, Croatia.}
\author{I.~Szentp\'{e}tery}
\affiliation{KFKI Research Institute for Particle and Nuclear
Physics,
             Budapest, Hungary.}
\author{J.~Sziklai}
\affiliation{KFKI Research Institute for Particle and Nuclear
Physics,
             Budapest, Hungary.}
\author{P.~Szymanski}
\affiliation{CERN, Geneva, Switzerland.}
\affiliation{Institute
for Nuclear Studies, Warsaw, Poland.}
\author{V.~Trubnikov}
\affiliation{Institute for Nuclear Studies, Warsaw, Poland.}

\author{D.~Varga}
\affiliation{KFKI Research Institute for Particle and Nuclear
Physics,
             Budapest, Hungary.}
\affiliation{CERN, Geneva, Switzerland.}
\author{M.~Vassiliou}
\affiliation{Department of Physics, University of Athens, Athens,
Greece.}
\author{G.I.~Veres}
\affiliation{KFKI Research Institute for Particle and Nuclear
Physics,
             Budapest, Hungary.}
\affiliation{MIT, Cambridge, USA.}
\author{G.~Vesztergombi}
\affiliation{KFKI Research Institute for Particle and Nuclear
Physics,
             Budapest, Hungary.}
\author{D.~Vrani\'{c}}
\affiliation{Gesellschaft f\"{u}r Schwerionenforschung (GSI),
             Darmstadt, Germany.}
\author{A.~Wetzler}
\affiliation{Fachbereich Physik der Universit\"{a}t, Frankfurt,
Germany.}
\author{Z.~W{\l}odarczyk}
\affiliation{Institute of Physics \'Swi\c etokrzyska Academy,
Kielce, Poland.}
\author{A.~Wojtaszek}
\affiliation{Institute of Physics \'Swi\c etokrzyska Academy,
Kielce, Poland.}
\author{I.K.~Yoo}
\affiliation{Department of Physics, Pusan National University,
Pusan,
             Republic of Korea.}
\author{J.~Zim\'{a}nyi}~\email[Deceased]{}
\affiliation{KFKI Research Institute for Particle and Nuclear
Physics,
             Budapest, Hungary.}


\collaboration{The NA49 collaboration} \noaffiliation



\begin{abstract}
The elliptic flow of $\Lambda$ hyperons has been measured by the
NA49 collaboration at the CERN-SPS in semi-central Pb+Pb
collisions at 158\agev. The standard method of correlating
particles with the event plane was used. Measurements of $\vi$
near mid-rapidity are reported as a function of rapidity,
centrality and transverse momentum. Elliptic flow of $\Lambda$
particles increases both with the impact parameter and with the
transverse momentum. It is compared with $\vi$ for pions and
protons as well as with model calculations. The observation of
significant elliptic flow and its mass dependence suggest strong
collective behavior of the matter produced in collisions of heavy
nuclei already at the SPS. Scaling properties of elliptic flow of
different particle species have been tested at 158$\agev$. The
limited $\pt$ range of the data does not allow for a decisive test
of the coalescence model.
\end{abstract}

\pacs{25.75.Ld}

\maketitle


\section{Introduction}
Elliptic flow in relativistic nuclear interactions has its origin
in the spatial anisotropy of the initial reaction volume in
non-central collisions and in particle rescatterings in the
evolving system which convert the spatial anisotropy into a
momentum anisotropy~\cite{Ollitrault:1997vz}. The spatial
anisotropy decreases rapidly because of the fast expansion of the
system~\cite{Kolb:2000sd} making the momentum anisotropy measured
at the end of this evolution strongly dependent on the matter
properties and the equation of state (EoS) at the early
stage~\cite{early_stages,Teaney:2000cw}. Comparison of measured
anisotropies with hydrodynamic model calculations provide an
important test of the degree of thermalisation in the produced
particle system at the early stage. Flow of heavy particles is
affected more strongly by changes in the EoS than flow of
pions~\cite{Teaney:2000cw,Huovinen:2001cy,Snellings:2003mh}.

The anisotropic flow parameters measured to date at SPS and lower
energies are mainly those of pions and
protons~\cite{v2_pi_p,Alt:2003ab}. RHIC experiments have measured
elliptic flow for many particle species~\cite{Adams:2004bi},
including hyperons~\cite{Adams:2004bi,Adams:2005zg}. In these
experiments, the rapid rise of elliptic flow with transverse
momentum ($\pt$) up to 1.5~\gevc~and its particle-mass dependence
are well reproduced by hydrodynamic models~\cite{Adams:2004bi}. At
higher values of $\pt$ quark number scaling of elliptic flow has
been observed at RHIC~\cite{Adams:2005zg,Adler:2003kt} which
indicates that the quark coalescence mechanism dominates hadron
production in the intermediate $\pt$
region~\cite{coalescence_models}. In order to test the validity of
this scenario at SPS energies we have extended elliptic flow
measurements in 158\agev~Pb+Pb ($\sqrt{s_{\rb{NN}}}=17.2$ GeV)
collisions to $\Lambda$ hyperons. In this paper, results on
$\Lambda$ elliptic flow as a function of the center of mass
rapidity ($y$) and transverse momentum will be presented and
compared to model calculations.

\section{Analysis}
The main components of the NA49 detector~\cite{NA49_setup} are
four large-volume Time Projection Chambers (TPCs) for tracking and
particle identification by energy loss ($dE$/$dx$) measurement
with a resolution of 3$-$6\%. The TPC system consists of two
vertex chambers inside the spectrometer magnets and two main
chambers placed behind the magnets at both sides of the beam.
Downstream of the TPCs a veto calorimeter detects projectile
spectators and is used for triggering and centrality selection.
The data sample consists of $3\times10^{6}$ semi-central Pb+Pb
events after online trigger selection of the 23.5\% most central
collisions. The events were divided into three different
centrality bins, which correspond to the first three bins used in
a previous analysis (see Table 1 in~\cite{Alt:2003ab}) and are
defined by centrality ranges 0-5\% (bin 1), 5-12.5\% (bin 2), and
12.5-23.5\% (bin 3). Many model predictions are published for
impact parameter ranges similar to those of our centrality
classes, which are: 0-3.4 fm (bin 1), 3.4-5.3 fm (bin 2), and
5.3-7.4 fm (bin 3) for our centrality classes. The measurement in
the centrality range $\sigma/\sigma_{\rb{TOT}}$ = 5$-$23.5\%
(called mid-central) is obtained by averaging the results of bins
2 plus 3 with weights corresponding to the fractions of the total
cross section in these bins.

The $\Lambda$ hyperon candidates were selected from the sample of
V$^0$-track configurations consisting of oppositely charged
particles, which include the $\Lambda$ decays into proton and
$\pi^-$ (branching ratio 63.9\%). The identification
method~\cite{Appelshauser:1999gs} relies on the evaluation of the
invariant mass distribution and is enhanced by daughter particle
identification applying a cut in $dE$/$dx$ around the expectation
value derived from a Bethe-Bloch parametrization. The extracted
$\Lambda$ candidates have a background contamination of 5-9\% in
the p-$\pi$ invariant mass window 1.108$-$1.124 \gevcc, depending
on centrality. The yields of $\Lambda$ hyperons are obtained by
counting the number of entries in the invariant mass peak above
the estimated background as a function of the azimuthal angle
$\phi$ of $\Lambda$ candidates with respect to the event plane
angle $\phiep$. The background is estimated from a fit of
invariant mass spectra to the sum of a Lorentz distribution and a
polynomial. It is subtracted in 16 bins of $\phi-\phiep$. The
acceptance of $\Lambda$ hyperons covers the range $0.4 \lesssim
\pt \lesssim 4$~\gevc~and $-1.5 \lesssim y \lesssim 1.0$ but the
detection efficiency strongly depends on azimuthal angle, $\pt$
and $y$. Thus we have to introduce differential corrections to
avoid biases when averaging $\vi$ over rapidity and transverse
momentum. Multiplicative factors were introduced for every
$\Lambda$ particle to correct the $\Lambda$ yields for detector
and reconstruction efficiency. They were determined as ratios of
published $\Lambda$ yields~\cite{Anticic:2003ux}, parametrized by
the Blast Wave model, to the measured raw $\Lambda$ yields.

The elliptic flow analysis is based on the standard procedure
outlined in~\cite{Alt:2003ab,Voloshin:1999gs} to reconstruct the
event plane for each event and the corrections for the event plane
resolution. The event plane is an experimental estimator of the
true reaction plane and is calculated from the azimuthal
distribution of primary charged $\pi$ mesons. Identification of
pions is based on $dE$/$dx$ measurements  in the TPCs. To avoid
possible auto-correlations, tracks associated with $\Lambda$
candidates are excluded from the event plane calculation. The
method to determine the event plane azimuthal angle $\phiep$ uses
the elliptic flow of pions, according to the formula:
\begin{eqnarray}
& X_{2}&= \sum_{i=1}^{N}p^{i}_{\rb{T}} [\cos(2\phi^{i})-\langle
\cos(2\phi)\rangle], \nonumber \\
& Y_{2}&= \sum_{i=1}^{N}p^{i}_{\rb{T}} [\sin(2\phi^{i})-\langle
\sin(2\phi)\rangle],  \\
& \phiep&= \frac{1}{2}
\tan^{-1}\left(\frac{Y_{\rb{2}}}{X_{\rb{2}}}\right), \nonumber
\end{eqnarray}
where $X_{\rb{2}},Y_{\rb{2}}$ are the components of the event
plane flow vector $\bf{Q_{\rb{{2}}}}$ and the sums run over
accepted charged pion tracks in an event. The acceptance
correction is based on the recentering method~\cite{Alt:2003ab}
which consists of subtracting in Eq.~1 the mean values $\langle
\cos(2\phi)\rangle$ and $\langle \sin(2\phi)\rangle$. These mean
values are calculated in bins of $\pt$ and rapidity for all pions
in those events which contain at least one $\Lambda$ hyperon
candidate. The values were stored in a 3-dimensional matrix of 20
$\pt$ intervals, 50 rapidity intervals, and eight centrality bins.
A second level acceptance correction is done by using mixed
events.
\begin{figure}[htb]
\begin{center}
\includegraphics[width=1.0\linewidth]{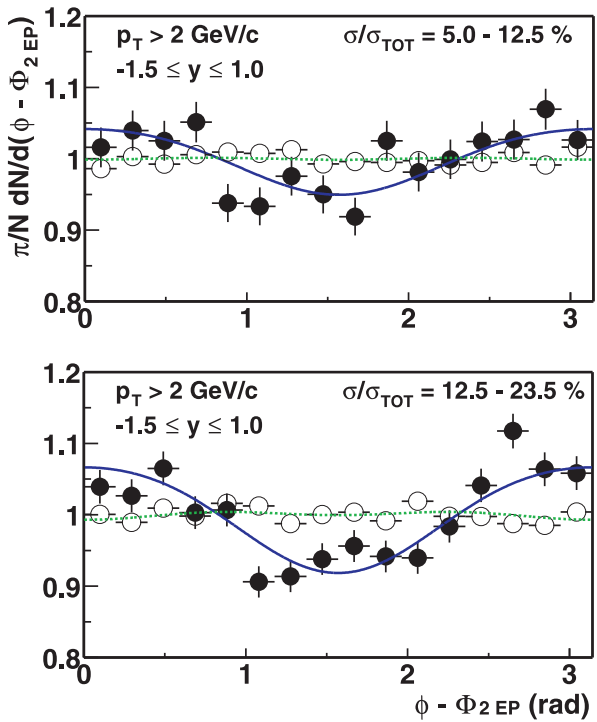}
\caption{(color online).~Azimuthal distributions of $\Lambda$
hyperons with respect to the event plane for real events (solid
symbols) and mixed-events (open symbols) in two centrality bins.
The curves are Fourier expansion fits (see text Eq. (2)).}
\end{center}
\end{figure}
We used 10 mixed events for each real event. Particles for mixed
events are randomly selected from different events in the same
centrality bin with at least one $\Lambda$ hyperon. The final
angular distributions are obtained by dividing the real $\Lambda$
angular distribution by the mixed event distribution to remove the
acceptance correlations remaining after recentering. The corrected
$\Lambda$ azimuthal distributions are then fitted with a truncated
Fourier series:
\begin{eqnarray}
\frac{dN}{d(\phi-\phiep)} = \nonumber \\
   \mbox{const}\times(1 &+& v^{obs}_{2}\cos[2(\phi-\phiep)]~~  \\
                 &+& v^{obs}_{4}\cos[4(\phi-\phiep)]).~~ \nonumber
\end{eqnarray}
 The elliptic flow $\vi$ is
evaluated by dividing the observed anisotropy $v^{obs}_{\rb{2}}$
by the event plane resolution $R$:
\begin{equation}
v_{\rb{2}}=\frac{v^{obs}_{\rb{2}}}{R}.
\end{equation}
The resolution,
\begin{equation}
R=\sqrt{2\langle
\cos[2(\Phi^{a}_{\rb{2EP}}-\Phi^{b}_{\rb{2EP}})]\rangle},
\nonumber
\end{equation}
is calculated from the correlation of two planes
($\Phi^{a}_{\rb{2EP}},\Phi^{b}_{\rb{2EP}}$) for random sub-events
with equal multiplicity. The results are $R$ = 0.27, 0.34 and 0.40
for centrality bins 1, 2 and 3, respectively. The total errors in
Figs.~2-4 are given by the quadratic sum of contributions from the
statistical error of the signal, the uncertainty due to the
background subtraction, the mixed event correction and event plane
resolution. The observed hexadecupole anisotropy
$v^{obs}_{\rb{4}}$ is consistent with zero within statistical
errors for all centrality bins.

\section{Results}
The final statistics in our sample consists of about $10^{6}$
$\Lambda$ candidates. This allows flow analysis for several
rapidity and $\pt$ bins.
\begin{figure}[htb]
\begin{center}
\includegraphics[width=1.0\linewidth]{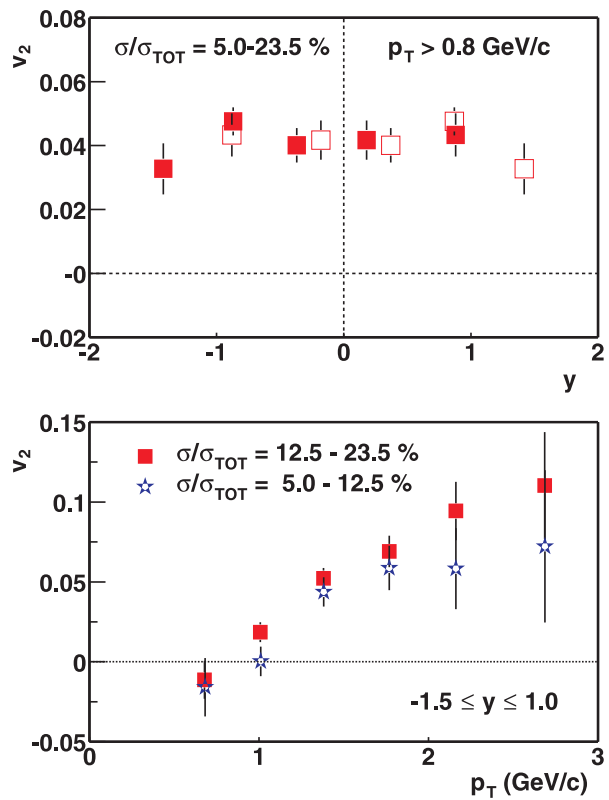}
\caption{(color online).~Elliptic flow of $\Lambda$ hyperons as a
function of rapidity (top) and $\pt$ (bottom). The open points in
the top graph have been reflected about midrapidity.}
\end{center}
\end{figure}
Two sample azimuthal distributions of $\Lambda$ hyperons with
respect to the estimated reaction plane for real and mixed events
are shown in Fig.~1. The curves represent results of fits with the
truncated Fourier series Eq.~(2). The distributions exhibit a
strong correlation for real events (full symbols and curves). As
expected, no correlation is observed for mixed-events (open
symbols, dashed curves). The $\pt$ averaged elliptic flow is
obtained from all identified $\Lambda$ hyperons without $\pt$
cuts. It exhibits no significant dependence on rapidity as shown
in Fig.~2(top). The absence
\vspace*{-0.0cm}
\begin{figure}[htb]
\begin{center}
\includegraphics[width=1.0\linewidth]{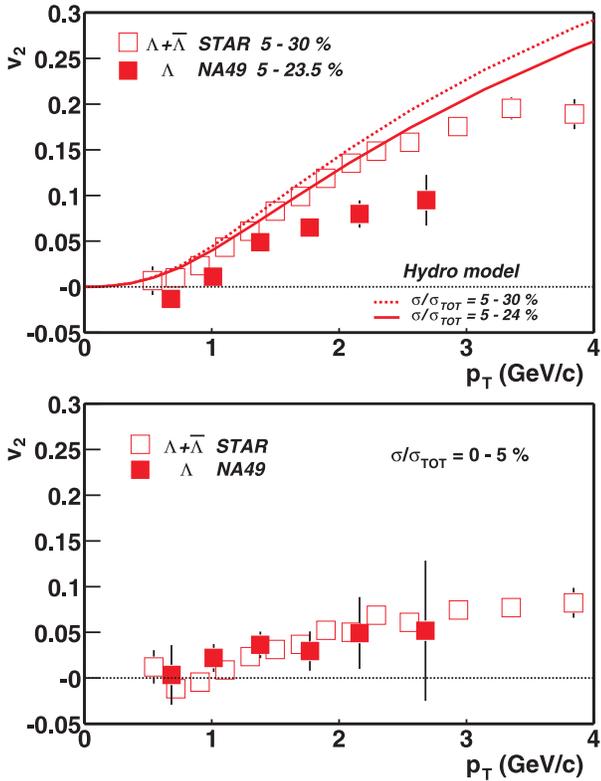}
\caption{(color online).~Elliptic flow of $\Lambda$ hyperons as a
function of $\pt$ from mid-central (top) and central (bottom)
events measured by the STAR (open symbols) and NA49 (solid
symbols) experiments. Curves are hydrodynamical model predictions
at RHIC energy for the two different centrality bins.}
\end{center}
\end{figure}
 of a rapidity dependence of $\vi(y)$ was also observed
for protons (see Fig.~6 in Ref.~\cite{Alt:2003ab}) in mid-central
events. We use the $\Lambda$ sample from the full rapidity range
of the data in Fig.~2(top) for the study of $\vi$ as a function of
$\pt$. The $\pt$ dependence of rapidity-averaged $\Lambda$
elliptic flow is shown in the bottom plot of Fig.~2 for two
centrality ranges. The $\vi$ parameter significantly increases
with transverse momentum, the rise being stronger for more
peripheral events. Fig.~3 shows a comparison of $\vi(\pt)$ of
$\Lambda$ hyperons for mid-central and central (0-5\%) events
measured by the NA49 and STAR experiments~\cite{Adams:2003am}. The
NA45 collaboration recently also presented preliminary results in
Pb+Au collisions at the top SPS energy~\cite{CERES_QM05} which
agree well with the NA49 results (not shown). For mid-central
collisions at SPS energy the elliptic flow grows linearly with
$\pt$ up to $\sim$2 GeV/c, but the increase is steeper at RHIC
than at SPS energy. It should be noted that RHIC mid-central data
have been measured in the centrality range
$\sigma/\sigma_{\rb{TOT}}$ = 5$-$30\% while SPS events are
somewhat more central. The effect
\begin{figure}[htb]
\begin{center}
\includegraphics[width=1.2\linewidth]{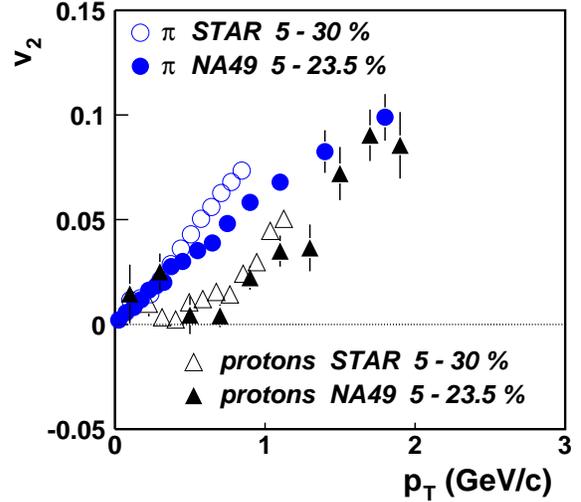}
\caption{(color online).~Elliptic flow for charged pions (circles)
and protons (triangles) as a function of $\pt$ from mid-central
events measured by the STAR~\cite{Adams:2004bi} (open symbols) and
NA49~\cite{Alt:2003ab} (solid symbols) experiments. }
\end{center}
\end{figure}
\begin{figure}[htb]
\begin{center}
\includegraphics[width=1.2\linewidth]{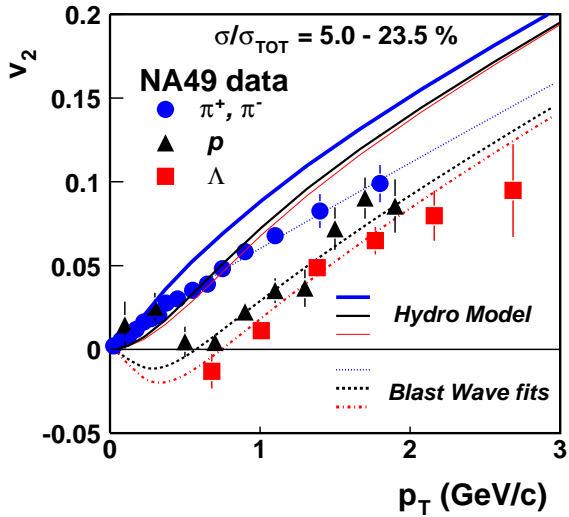}
\caption{(color online).~Elliptic flow for charged pions
(circles), protons (triangles) and $\Lambda$ hyperons (squares) as
a function of $\pt$ from 158\agev~Pb+Pb mid-central events
measured by the NA49 experiment. Curves are blast wave fits and
hydrodynamic model predictions for $\sqrt{s_{\rb{NN}}}=17.2$ GeV.
}
\end{center}
\end{figure}
 of different centrality ranges has been estimated by
hydrodynamic calculations~\cite{Huovinen:2005gy,RHIC_hydro_params}
at RHIC energy for the slightly different centrality bins of NA49
and STAR. As shown by the corresponding curves in Fig.~3 this
explains only partially the difference between both measurements.
For central collisions NA49 and STAR results do not differ
significantly (Fig.~3 bottom). A comparison of $\vi$ of pions and
protons obtained by NA49~\cite{Alt:2003ab} and
STAR~\cite{Adams:2004bi} for mid-central collisions is shown in
Fig.~4. The NA49 values were obtained as the cross section
weighted averages of the measurements published
in~\cite{Alt:2003ab} for the appropriate centrality range. As
observed for $\Lambda$ hyperons, $\vi$ of pions also rises faster
with $\pt$ at RHIC than at SPS. Protons seem to exhibit a similar
trend, but the limited $\pt$ range and the larger measurement
errors do not allow a firm conclusion.

A comparison of $\vi(\pt)$ for pions, protons and $\Lambda$
hyperons as measured by the NA49 experiment in mid-central events
is displayed in Fig.~5. The elliptic flow grows linearly with
$\pt$ for all particle species but the rise for pions starts from
$\pt$ close to zero while for protons and $\Lambda$ particles it
starts from $\pt\approx0.5$~\gevc. The elliptic flow for pions is
significantly larger than that for heavier particles although at
$\pt \approx 2$~\gevc~the flow becomes similar for all particle
species. Data are reproduced by blast wave
fits~\cite{Huovinen:2001cy,Adler:2001nb} (dashed curves in Fig.~5)
with the following parameters: freeze-out temperature $T$~=~92
MeV, mean transverse expansion rapidity of the shell
$\rho_{0}$~=~0.82, its second harmonic azimuthal modulation
amplitude $\rho_{a}$~=~0.021, and a spatial eccentricity parameter
$s_{2}$~=~0.033. The values of freeze-out temperature and
expansion rapidity are consistent with those obtained from $\mt$
spectra and Bose-Einstein correlations~\cite{NA49_HBT}. In Fig.~5
the measured values of $\vi$ are also compared to hydrodynamical
model calculations~\cite{Huovinen_private} assuming a first-order
phase transition to a QGP at the critical temperature
$T_{\rb{c}}=165$ MeV. The initial conditions employed for the
hydrodynamical calculations at SPS are: initial energy density
$\epsilon_{0}$~=~9.0~$\gevf$, baryon density $n_{b}$~=~1.1~$\fmm$,
thermalization time $\tau_{0}$~=~0.8~$\fmc$. The initial
conditions were fixed as in~\cite{Hydro_conditions} by requiring a
good fit to the $\pt$ spectra of protons and negatively charged
particles in central Pb+Pb collisions at the SPS. With the
freeze-out temperature $T_{\rb{f}}=120$ MeV tuned to reproduce
particle spectra, the model calculations significantly
overestimate the SPS results for semi-central collisions (full
curves in Fig.~5) in contrast to predictions at RHIC energy which
agree quite well with data for $\pt \lesssim
2$~\gevc~\cite{Adams:2003am} (see Fig.3). The discrepancy at SPS
may indicate a lack of complete thermalisation or a viscosity
effect. However, the model reproduces qualitatively the
characteristic hadron-mass ordering of elliptic flow. Thus the
data support the hypothesis of early development of collectivity.
On the other hand, the magnitude and trends of elliptic flow,
measured by the second Fourier coefficient $\vi$ is found to be
underpredicted by a hadronic cascade model~\cite{Bleicher}. A more
comprehensive description of experimental results was found by
coupling a hadronic rescattering phase to the hydrodynamical
evolution and hadronisation~\cite{Teaney:2000cw}. This model can
reproduce $\mt$ spectra and elliptic flow both at top SPS energy
and RHIC consistently, although predictions of $\vi$ for exactly
the centrality range  of the present analysis are not published in
the literature. One should note, however, that none of the
hydrodynamical models have yet been able to describe Bose-Einstein
correlations successfully.

Quark coalescence models~\cite{coalescence_models} have been used
to explain the quark number scaling observed at RHIC for $\vi$ in
the intermediate $\pt$ region.
\begin{figure}[htb]
\begin{center}
\includegraphics[width=0.93\linewidth]{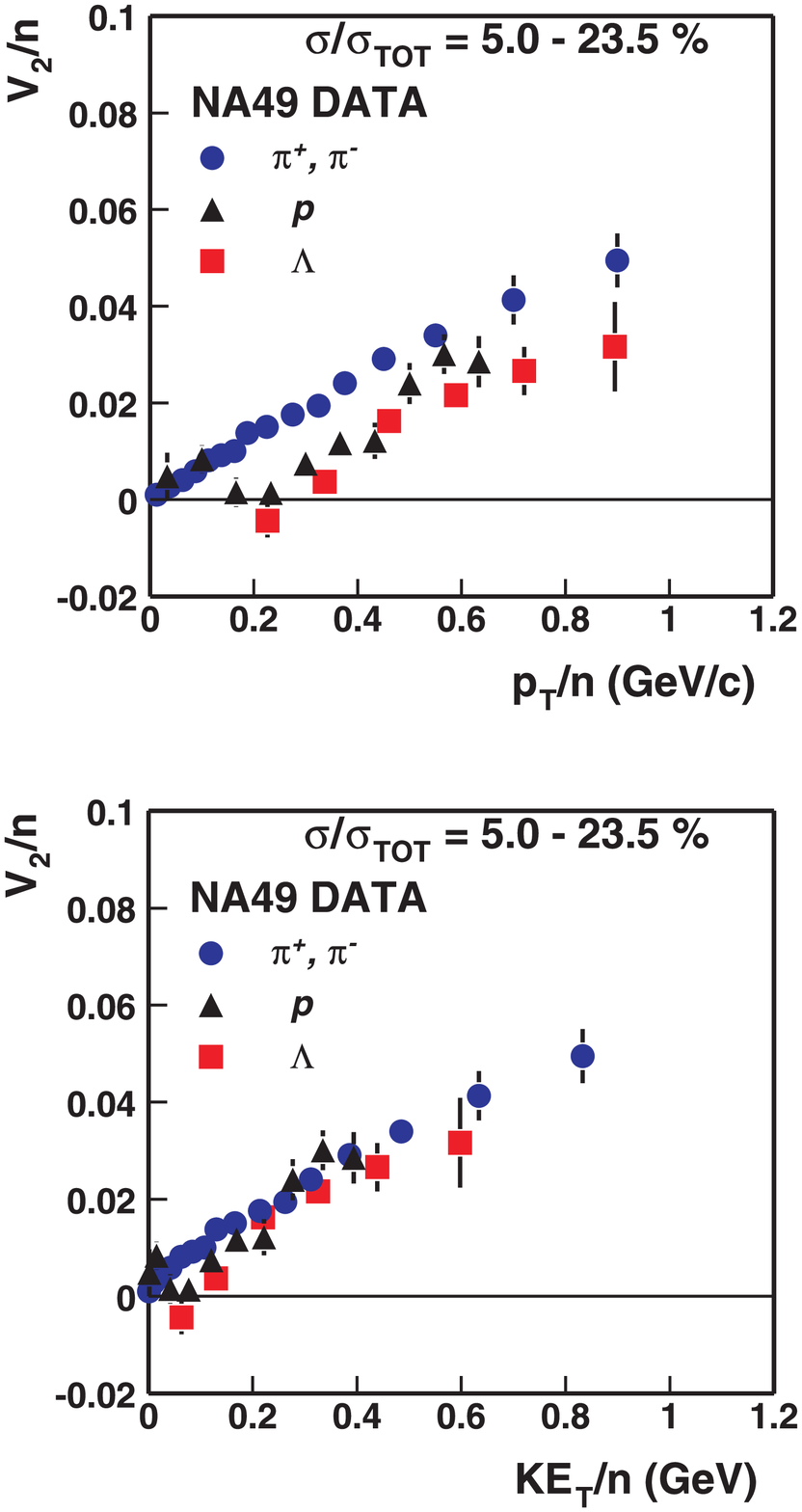}
\caption{(color online).~Elliptic flow for charged pions
(circles), protons (triangles) and $\Lambda$ hyperons (squares)
scaled by the number of constituent quarks $\vi/n$ as a function
of $\pt/n$ (top) and $\ket/n$ (bottom) from 158\agev~Pb+Pb
mid-central events measured by the NA49 experiment. }
\end{center}
\end{figure}
Fig.~6 shows the scaling behavior of the $\vi$ measurements of
NA49 at the SPS. The values of $\vi$ shown in Fig.~5 were divided
by the number of constituent quarks $n=3$ for baryons and $n=2$
for mesons.

When plotting $\vi/n$ versus $\pt/n$ approximate scaling was
observed at RHIC~\cite{Adams:2004bi,Adler:2003kt} except for
pions. The NA49 measurements (Fig.~6, top) are consistent with
this result in the $\pt$ range covered by the data. The deviation
of the pions from the universal curve has been attributed to the
Goldstone nature of the pion (its mass is smaller than the sum of
masses of its constituent quarks) or to the effect of resonance
decays~\cite{pion_coales}.

The variable $\ket=\mt-m$, where $\mt$ is the transverse mass and
$m$ the rest mass of the particle, was proposed
in~\cite{ket_scaling} as an alternative to $\pt$ since pressure
gradients which give rise to azimuthal asymmetry may naturally
lead to collective transverse energy of produced particles. Good
scaling for all particle species is seen when $\vi/n$ is plotted
versus $\ket/n$ in the range $\ket/n \lesssim 0.8$~GeV covered by
the SPS data (Fig.~6, bottom). This behaviour was first observed
at RHIC and interpreted as a result of hydrodynamic
evolution~\cite{ket_scaling}. The data of NA49 do not reach higher
transverse momenta at which a decisive test of coalescence models
would be possible.


\section{Summary}
In summary, we report the first measurement of the anisotropic
flow parameter $\vi$ for $\Lambda$ particles from Pb+Pb collisions
at $\sqrt{s_{\rb{NN}}}=17.2$ GeV. Elliptic flow of $\Lambda$
hyperons exhibits no significant dependence on rapidity for $-1.5
\lesssim y \lesssim 1.0$. It rises linearly with $\pt$ and is
smaller than $\vi$ for pions. Both features are quantitatively
reproduced by the Blast Wave parametrization but only
qualitatively by the hydrodynamic model. The increase of $\vi$
with $\pt$ is weaker at SPS than at RHIC energy. The observation
of significant elliptic flow and its mass dependence suggest
strong collective behaviour of the matter produced in collisions
of heavy nuclei already at the SPS. Hydrodynamic models with a
deconfinement phase transition and a microscopic freezeout
treatment appear to provide a consistent description of $\vi$ and
$\mt$ spectra at both top SPS and RHIC
energies~\cite{Teaney:2000cw}. Quark number scaling of elliptic
flow ($\vi/n$ versus $\ket/n$) was shown to hold also at the SPS.
However, the limited $\pt$ reach of the data does not allow a
decisive test of the quark coalescence hypothesis.


\begin{acknowledgements}

This work was supported by the US Department of Energy Grant
DE-FG03-97ER41020/A000, the Bundesministerium f\"{u}r Bildung und
Forschung Grant 06F-137, Germany, the Virtual Institute VI-146 of
Helmholtz Gemeinschaft, Germany, the Polish State Committee for
Scientific Research (1 P03B 006 30, 1 P03B 097 29, 1 P03B 121 29,
1 P03B 127 30), the Hungarian Scientific Research Foundation
(T032648, T032293, T043514), the Hungarian National Science
Foundation, OTKA, (F034707), the Polish-German Foundation, the
Korea Science \& Engineering Foundation (R01-2005-000-10334-0) and
the Bulgarian National Science Fund (Ph-09/05).
\end{acknowledgements}



\begin{thebibliography}{9}
\bibitem{RHIC_hydro_params} The initial conditions employed for the hydrodynamical calculations at RHIC are:
$\epsilon_{0}$~=~23~$\gevf$, $n_{b}$~=~0.12~$\fmm$,
$\tau_{0}$~=~0.6~$\fmc$ .
\end{thebibliography}
\end{document}